\documentclass[twocolumn]{aastex631}
\usepackage{graphicx}
\usepackage{amsmath,amssymb}

\shorttitle{Ion-scale transition of plasma turbulence}
\shortauthors{Hellinger et al.}

\begin{document}

\title{Ion-scale transition of plasma turbulence:
Pressure-strain effect}

\author[0000-0002-5608-0834]{Petr Hellinger}
\affiliation{Astronomical Institute, CAS,
Bocni II/1401,CZ-14100 Prague, Czech Republic}
\affiliation{Institute of Atmospheric Physics, CAS,
Bocni II/1401, CZ-14100 Prague, Czech Republic}
\email{petr.hellinger@asu.cas.cz}
\author[0000-0002-7848-9200]{Victor Montagud-Camps}
\affiliation{Astronomical Institute, CAS,
Bocni II/1401,CZ-14100 Prague, Czech Republic}
\author[0000-0002-7419-0527]{Luca Franci}
\affiliation{Queen Mary University of London, UK}
\author[0000-0002-6276-7771]{Lorenzo Matteini}
\affiliation{Imperial College, London, UK}
\author[0000-0002-7969-7415]{Emanuele Papini}
\affiliation{INAF -- Istituto di Astrofisica e Planetologia Spaziali, via del Fosso del Cavaliere 100, I-00133 Roma, Italy}
\author[0000-0003-4380-4837]{Andrea Verdini}
\affiliation{Dipartimento di Fisica e Astronomia, Universit\`a degli Studi di Firenze Largo E. Fermi 2, I-50125 Firenze, Italy}
\affiliation{INAF -- Osservatorio Astrofisico di Arcetri, Largo E. Fermi 5, I-50125 Firenze, Italy}
\author[0000-0002-1322-8712]{Simone Landi}
\affiliation{Dipartimento di Fisica e Astronomia, Universit\`a degli Studi di Firenze Largo E. Fermi 2, I-50125 Firenze, Italy}
\affiliation{INAF -- Osservatorio Astrofisico di Arcetri, Largo E. Fermi 5, I-50125 Firenze, Italy}

\begin{abstract}

We investigate properties of solar wind-like plasma turbulence using
direct numerical simulations. We analyze the transition
from large, magnetohydrodynamic (MHD) scales to the ion characteristic ones
using two-dimensional hybrid (fluid electrons, kinetic ions) simulations.
To capture and quantify turbulence properties, we apply the Karman-Howarth-Monin (KHM)
equation for compressible Hall MHD (extended by considering
the plasma pressure as a tensor quantity) to the numerical results.
The KHM analysis indicates that the transition from MHD to ion scales (the
so called ion break in the power spectrum) results from a combination of an onset of Hall physics and
of an effective dissipation owing to the pressure-strain energy-exchange channel
and resistivity. We discuss the simulation results in the context of the solar wind.

\end{abstract}

\keywords{methods: numerical  --- plasmas  --- solar wind  --- turbulence}

\section{Introduction} \label{intro}

Turbulence in the weakly-compressible solar wind plasma
exhibits a clear transition at ion scales \citep{brca13}.
At large scales, magnetic power spectra of time series observed in situ
have typically a power-law dependence on the frequency
with a spectral index close to the Kolmogorov $-5/3$
phenomenological prediction for hydrodynamic turbulence. 
Around scales corresponding to ion characteristic scales (the ion gyroradius and inertial length) the power spectra
steepen. This steepening was initially regarded as a signature of
the dissipation onset \citep{leamal98}. However, the Hall term starts to
play on similar scales and leads also to a spectral
steepening without necessarily implying the presence of some energy dissipation \citep{ghosal96,galt06,papial19}.
 
One way how to discern and quantify different turbulent
processes is the 
K\'arm\'an-Howarth-Monin (KHM) equation 
\citep{kaho38,moya75,fris95} that connects 
the energy decay/injection with its cascade and
dissipation.
Recently, the 
incompressible version of the KHM equation for the  Hall MHD
\citep{popo98b,galt08,hellal18,ferral19}
was used to study the ion transition
in simulations as well as observations
\citep{hellal18,bandal20b,adhial21}.
These results indicate that at ion
scales there is indeed a transition
from a MHD to a Hall dominated turbulent
cascade. However, through this transition 
the total cascade rate is observed to decrease and this
suggests some sort of dissipation. Consequently,
based on the previous works above,
the ion transition is likely a combination
of the Hall physics onset and dissipation.

Theoretical analyses \cite[][and references therein]{yangal17}
shows that one possible channel that
leads to energy exchanges between the
magnetic + particle kinetic energy and
the particle internal energies in collisionless plasmas is the
pressure-strain effect, a generalization of
the pressure-dilation effect. Results of kinetic simulations
 \citep{yangal19,mattal20} indicate that
this effect can act as an effective dissipation 
and may explain the decrease of the energy cascade rate observed
at ion scales.
 Moreover, numerical simulations and in situ observations
 suggest that the pressure-strain channel is likely responsible for 
the correlations between particle velocity-field gradients and temperatures
\citep{franal16a,pama16,yangal19,pezzal21,yordal21}.

{
In this paper we revisit the work of \cite{hellal18} and analyze
the kinetic simulation results using both the incompressible and compressible versions of the KHM
equation in order to test the validity of these approximations.
We use the compressible KHM equation derived by \cite{hellal21b}
because, motivated by the previous works of \cite{yangal17} and \cite{mattal20}, we want to determine effects of the pressure-strain coupling
while the alternative approaches \cite[][and references therein]{andral18}
assume a scalar pressure along with 
some particular closure, thus preventing
their extension to the weakly collisional case
with a tensor description of the particle pressure.

This paper is organized as follows: In section~\ref{simul}
we present spectral properties of three two-dimensional hybrid
simulation. In section~\ref{iKHM} we analyze the hybrid
simulations using the incompressible KHM equation.
In section~\ref{cKHM} we extend the analysis to the
compressible KHM equation. In section~\ref{discus} we
summarize and discuss the presented results. 
}

\section{Simulation results}
\label{simul}
Here we analyze two-dimensional (2D)
hybrid simulations of decaying plasma turbulence using the KHM equations.
 In the hybrid approximation, ions are
described by a particle-in-cell model whereas electrons are a massless, charge
neutralizing fluid \citep{matt94}. 
We use the 2D version of the code 
Camelia (\url{http://www.asu.cas.cz/\textasciitilde helinger/camelia.html})
using a simulation setup that is similar
to that of \cite{franal15b}. 

\begin{deluxetable}{cccccccc}
\tablenum{1}
\tablecaption{List of simulations and their relevant parameters.}
\label{tabsim}
\tablehead{
\colhead{Run} &
\colhead{ $\beta_i$} &
\colhead{ $\delta B/B_0$} &
\colhead{ $k_{\mathrm{inj}} d_i$}  &
\colhead{$\Delta x /d_i$} &
\colhead{ $\eta$} &
\colhead{ $N_{ppc}$} &
\colhead{ $\Delta t\Omega_i$}
}
\startdata
1  &$0.1$ & $0.25$ & $0.22$  & $1/16$ & $4\times 10^{-4}$  & $8192$  & $0.005$ \\
2  & $0.5$ & $0.25$ & $0.22$  & $1/16$ & $4\times 10^{-4}$ & $8192$ &$0.005$ \\
3  & $2.5$ & $0.25$ & $0.2$  & $1/8$ & $4\times 10^{-4}$ & $8192$ &$0.01$ \\
\enddata
\tablecomments{The resistivity $\eta$ is given in the units of $\mu_0 v_{A}^2/\Omega_{i}$}
\end{deluxetable}
{
We investigate properties of three hybrid simulations with parameters similar
(but not identical) to those in \cite{hellal18}; in contrast to this work, 
we try here to use the same/similar parameters for all the runs if possible, see Table~\ref{tabsim}.}
Protons are initially isotropic with different values of $\beta_{i}$ (run~1: $0.1$,
run~2: $0.5$, and run~3: $2.5$). We consider a 2D domain $(x,y)$ of size $2048
\times 2048$ grid points and resolution $\Delta x = \Delta y =d_i/16$ (for runs~1 and~2)
and $d_i/8$ (for run~3). Here $d_i$ denotes the ion inertial length and $\beta_{i}$
stands for the ion beta, i.e., the ratio between the ion and magnetic pressures.
 In order to reduce the noise, a Gaussian smoothing on $3\times 3$ points is used
on the proton density and velocity in the code.
A uniform ambient
magnetic field $\boldsymbol{B}_0$, directed along $z$ and perpendicular to the
simulation domain is present whereas neutralizing electrons are assumed
isotropic and isothermal. Furthermore, we set the electron beta  
(i.e., the ratio between electron and magnetic pressures) equal 
to the initial value of the ion beta ($\beta_{e}=\beta_{i}$).
 The system is perturbed
with an isotropic 2-D spectrum of modes with random phases, linear Alfv\'en
polarization ($\delta \boldsymbol{B} \perp \boldsymbol{B}_0$) and vanishing
correlation between magnetic field and velocity fluctuations.  These modes are
in the range  $k\le 0.22 d_i^{-1}$ (for runs 1 and 2) and $k\le 0.2 d_i^{-1}$ (for run~3)  and have a flat
one-dimensional power spectrum with rms fluctuations $\delta B=0.25$. The time step is
$\Delta t=0.005 \Omega_i^{-1}$  (for runs 1 and 2) and $0.01 \Omega_i^{-1}$ (for run~3)  
for particles integration (the magnetic field is advanced with a
smaller time step $\Delta t_B = \Delta t/20$), the number of particle per cell
$N_{ppc}=8192$, and a small resistivity $\eta=4\times 10^{-4} \mu_0 v_{A}^2/\Omega_{i}$ is used to avoid energy accumulation at the
smallest scales {(note that no explicit viscosity is present in the hybrid model)}. 
Here $\Omega_{i}$ denotes the ion cyclotron frequency, $\mu_0$ is the magnetic permeability of the vacuum,
and $v_{A}$ stands for the Alfv\'en velocity.
We let the system evolve beyond the
time when it becomes quasi-stationary \citep{mipo09,serval15};
henceforth, we analyze properties of plasma turbulence at such times 
$t_d=286\Omega_i^{-1}$, $290\Omega_i^{-1}$, and $299\Omega_i^{-1}$
for run~1, 2, and 3, respectively.

\begin{figure}[htb] 
\centerline{\includegraphics[width=7.5cm]{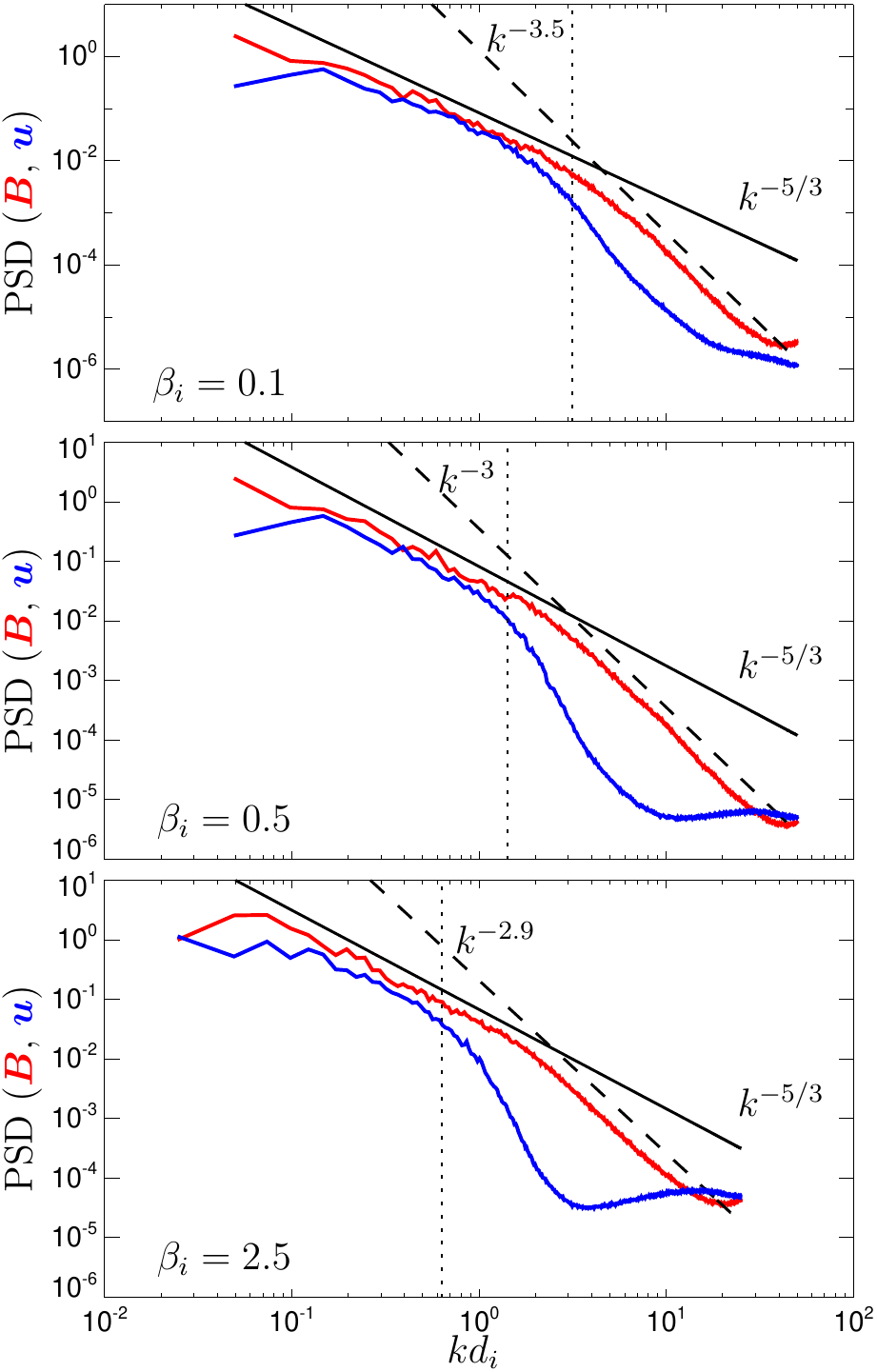}}
\caption{Power spectral densities of (red) the magnetic field and (blue) the
proton velocity field in the three simulations. 
The dotted lines denote $k\rho_i=1$.
 \label{specb}}
\end{figure}

Figure~\ref{specb} shows the power spectral density of the
magnetic field for the three simulations.  The simulated spectra exhibit two
power laws with a smooth transition at ion scales (the so called ion spectral
break), whose shape and position depend on the plasma beta: its scale
is close to $d_i$ for small betas whereas in high beta plasmas its around $\rho_i$ 
  \citep{franal16b}. Here $\rho_i$ denotes the ion gyroradius. 
The magnetic power spectral slopes on large scales is about $-5/3$,
in the sub-ion range the  
spectrum steepens, with slopes about $-3.5$, $-3$, $-2.9$ in the three simulations
\cite[cf.,][]{franal16b}. On the other hand, the magnetic compressibility increases at ion scales
depending on the plasma beta \citep{mattal20b}.
The proton velocity fluctuating field
has a power spectrum on large scales with a similar slope and decouples from
the magnetic fluctuations around the proton gyroscales. The sub-ion velocity
fluctuations have a limited scale range before reaching the noise level so that
it is difficult to distinguish between an exponential and a power-law
dependence in the sub-ion range.

\section{Incompressible KHM Analysis}
\label{iKHM}
We start with the incompressible, constant-density,
inviscid {(there is no explicit viscosity in the hybrid model)} but resistive Hall MHD approximation.
In this case we have the energy (per unit mass) budget equation for
the kinetic+magnetic energy 
\begin{align}
\frac{\partial}{\partial t}\left\langle \frac{1}{2} \left(|\boldsymbol{u}|^2+|\boldsymbol{b}|^2\right)  \right\rangle = \epsilon
\end{align}
where 
$\boldsymbol{u}$ is the velocity field, $\boldsymbol{b}$ is
the magnetic field in the Alfv\'en units ($\boldsymbol{b}=\boldsymbol{B}/\rho^{1/2}$,
$\boldsymbol{B}$ being the magnetic field and $\rho$ the plasma
density assumed to be constant; here we assume SI units except for the magnetic permeability $\mu_0$ that is set to one)
and
\begin{align}
\epsilon=\eta \langle  \boldsymbol{\nabla} \boldsymbol{b}: \boldsymbol{\nabla} \boldsymbol{b} \rangle
\end{align}
is the resistive dissipation rate (per unit mass).
Following \cite{hellal18} we define the effective
dissipation/cascade rate (per unit mass) $\epsilon^*$ given by the corresponding KHM equation (the superscript
$(i)$ will henceforth denote incompressible, constant-density quantities)
\begin{equation} 
\epsilon^*= -\frac{1}{4}\frac{\partial S^{(i)}}{\partial t}
+ K_\text{MHD}^{(i)} + K_\text{H}^{(i)}
+\frac{\eta}{2}\Delta S_{b}^{(i)} 
\label{epsilonstar}
\end{equation}
where $S_{b}^{(i)}$, $S_{u}^{(i)}$, and $S^{(i)}=S_{b}^{(i)}+S_{u}^{(i)}$
are  the second-order structure functions
\begin{align}
S_{b}^{(i)}=\langle\left|\delta\boldsymbol{b}\right|^{2}\rangle, \ \ \
S_{u}^{(i)}=\langle\left|\delta\boldsymbol{u}\right|^{2}\rangle,
\end{align}
and $K_\text{MHD}^{(i)}$ and $K_\text{H}^{(i)}$  are given by
\begin{align}
K_\text{MHD}^{(i)}=-\frac{1}{4} \boldsymbol{\nabla}\cdot \boldsymbol{Y}^{(i)}, \ \ \text{and} \ \
K_\text{H}^{(i)}=- \frac{1}{4} \boldsymbol{\nabla}\cdot \boldsymbol{H}^{(i)},
\end{align}
respectively, involving the third structure functions
\begin{align}
  \boldsymbol{Y}^{(i)}&=\left\langle
  \delta\boldsymbol{u}\left(\left|\delta\boldsymbol{u}\right|^{2}+\left|\delta\boldsymbol{b}\right|^{2}\right)-2\delta\boldsymbol{b}\left(\delta\boldsymbol{u}\cdot\delta\boldsymbol{b}\right)\right\rangle, \\
\boldsymbol{H}^{(i)}&=\left\langle  \delta\boldsymbol{b}\left(\delta\boldsymbol{b}\cdot\delta\boldsymbol{j}\right) - \frac{1}{2} \delta\boldsymbol{j}\left|\delta\boldsymbol{b}\right|^{2}\right\rangle.
\end{align}
Here $\delta$ denotes the increment for the spatial separation $\boldsymbol{l}$,
$\delta\boldsymbol{u}=\boldsymbol{u}(\boldsymbol{x}+\boldsymbol{l})-\boldsymbol{u}(\boldsymbol{x})$,
etc., $\boldsymbol{j}$ is the electric current in the Alfv\'en units 
($\boldsymbol{j}=\boldsymbol{J}/(e n)$, $\boldsymbol{J}$ being the electric current
$e$ the proton charge, $n$ the proton number density).
In Equation~(\ref{epsilonstar}), $S_{u}^{(i)}$ and $S_{b}^{(i)}$ represent 
the separation-scale distribution of the kinetic and magnetic energy (per unit mass), respectively,
$K_\text{MHD}^{(i)}$ and $K_\text{H}^{(i)}$ are the MHD and Hall cascade rates,
respectively, and $\eta\Delta S_{b}^{(i)}/2$ describes the resistive dissipation.

\begin{figure}[thb]
\centerline{\includegraphics[width=7.5cm]{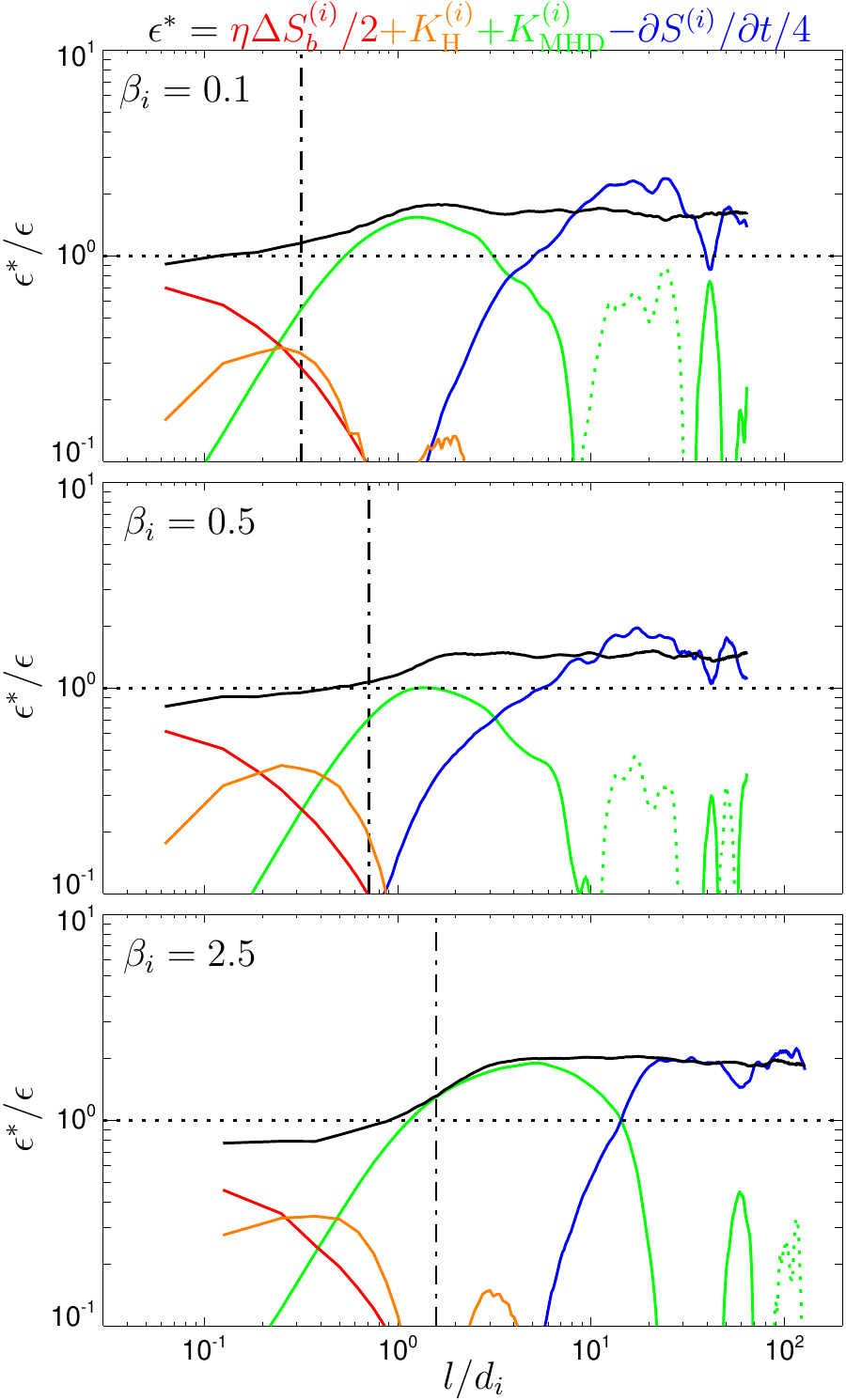}}
\caption{Incompressible KHM Hall MHD equation results for 
the three simulation (from top to bottom:
$\beta_i=0.1$, $\beta_i=0.5$, and
$\beta_i=2.5$): The cascade rate $\epsilon^*$ normalized to the 
resistive heating
rate $\epsilon$ as a function of the spatial scale separation $l$ is shown as a black curve. The different
contributing terms to $\epsilon^*$ are also shown as: (blue) $-{\partial
S^{(i)}}/{\partial t}/4$, (green) $K_\text{MHD}^{(i)}$,
(orange) $K_\text{H}^{(i)}$, and (red) $\eta\Delta
S_{b}^{(i)}/2$ (note that dotted lines denote negative values).
  The dash-dotted
lines denote $l=\rho_i$. \label{iyag}}
\end{figure}

Equation~(\ref{epsilonstar}) constitutes an energy-per-mass budget relationship,
it describes the balance between injection, energy cascade flux and dissipation;
 the KHM equation
corresponds to the case when $\epsilon^*=\epsilon=\text{const}$. 
We can now directly verify the predicted constancy of the right-hand side of 
Eq.~(\ref{epsilonstar}) in the simulations, and compare the relative importance 
of the different terms across the separation scales.
Figure~\ref{iyag} shows the results of such analysis, displaying
the effective cascade rate $\epsilon^*$ and its contributing terms as functions of
the scale separation $l$ for the three simulations. 
The different terms are calculated at the times $t_d$; 
$\partial S/\partial t$ is estimated by the finite difference 
$(S(t_d+\Delta t)-S(t_d))/\Delta t$ using the simulation (particle) time step.

Figure~\ref{iyag} demonstrates that the effective cascade rate
in the three simulations $\epsilon^*$ is constant on large scales but larger
than the expected resistive dissipation rate $\epsilon$.
The cascade rate $\epsilon^*$ decreases at ion scales and this transition
shifts to larger scales for larger $\beta_i$ (and larger $\rho_i$). 
{This behavior was already seen in \cite{hellal18}
but in this work the decrease was underestimated since 
the Hall terms had twice the correct value \cite[cf.,][]{ferral19}.}
These properties indicate the existence of an additional dissipation channel
 not captured in this analysis.

\section{Compressible KHM analysis}

\label{cKHM}
\begin{figure}[thb]
\centerline{\includegraphics[width=7.5cm]{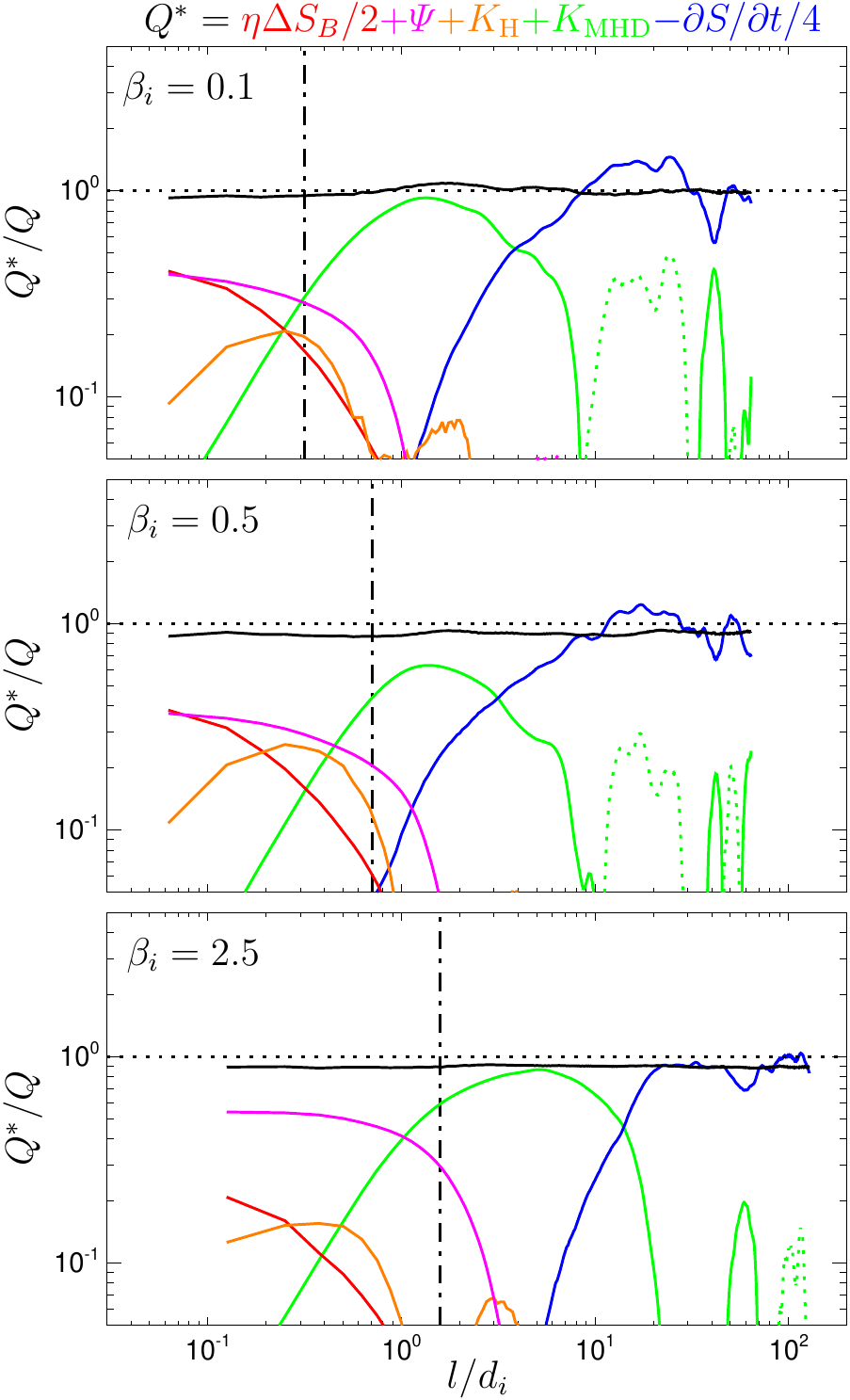}}
\caption{KHM equation results for $\beta_i=0.1$, $\beta_i=0.5$, and
$\beta_i=2.5$ (from top to bottom): The cascade rate $Q^*$ normalized to the total effective heating
rate $Q$ as a function of $l$ is shown as a black curve. The different
contributing terms to $\epsilon^*$ are also shown as: (blue) $-{\partial
S}/{\partial t}/4$, (green) $K_\text{MHD}$,
(orange) $K_\text{H}$, (magenta) $\varPsi$, and (red) $\eta\Delta
S_{b}/2$ (note that dotted lines denote negative values).
  The dash-dotted
lines denote $l=\rho_i$. \label{yag}}
\end{figure}
We now relax the incompressibility and constant-density assumption {(but we still neglect
the viscosity)}
and, moreover, we consider a weakly collisional plasma
where the pressure is described as a pressure tensor ($\boldsymbol{\mathrm{P}}$).
The kinetic+magnetic energy  budget equation reads then
\begin{align}
\frac{\partial}{\partial t}\left\langle \frac{1}{2} \left(|\boldsymbol{w}|^2+|\boldsymbol{B}|^2\right)  \right\rangle = -Q
\end{align}
where we include the (square root of the) plasma density into 
the velocity field $\boldsymbol{w}=\sqrt{\rho}\boldsymbol{u}$ \citep{kior90},
and where we define the total effective dissipation rate 
{(per unit volume here in contrast to the incompressible
case)}
\begin{align}
Q=Q_\eta+Q_\Sigma
\end{align}
with
\begin{align}
Q_\eta= \eta \langle \boldsymbol{\nabla} \boldsymbol{B}: \boldsymbol{\nabla}\boldsymbol{B} \rangle = \eta \langle J^2 \rangle
\end{align}
being the resistive dissipation rate {(simply related to the resistive dissipation rate per unit mass as 
$Q_\eta=\langle\rho\rangle \epsilon$; 
here $\langle\rho\rangle$ is the background/average plasma density needed 
to relate the normalizations per unit mass and per unit volume)} and
\begin{align}
Q_\Sigma=-\langle \boldsymbol{\mathrm{P}} : \boldsymbol{\Sigma} \rangle
\end{align}
representing the pressure-strain coupling \citep{yangal17} ($\boldsymbol{\Sigma}$ being the
velocity strain tensor, $\boldsymbol{\Sigma}=\boldsymbol{\nabla}\boldsymbol{u}$). 
To generalize Equation~(\ref{epsilonstar}),
we consider the compressible KHM equation \cite[Eq.~(4) of][where we neglect the viscous term]{hellal21b}.
We express the resistive dissipation rate as a function of the remaining terms
and add the pressure-strain term $Q_\Sigma$ to both sides and we define an effective cascade rate $Q^*$ as
\begin{align}
Q^*=-\frac{1}{4}\frac{\partial {S}}{\partial t} +{K}_\text{MHD} 
+{K}_\text{H} + \varPsi + \frac{\eta}{2} \Delta {S}_B.
\label{qstar}
\end{align}
Here the second order structure functions are given
\begin{align*}
{S}_{w} &=\left\langle |\delta\boldsymbol{w}|^2\right\rangle, \ \
{S}_{B}=\left\langle |\delta\boldsymbol{B}|^{2}\right\rangle
, \ \ \text{and}   \ \
{S}={S}_{w}+{S}_{B},
\end{align*}
and the following quantities,
\begin{align}
{K}_\text{MHD}&=-\frac{1}{4}\boldsymbol{\nabla}\cdot\boldsymbol{{Y}} -\frac{1}{4} {R} \nonumber
+\frac{1}{2} C_{\sqrt{\rho}}\left[\boldsymbol{u},\boldsymbol{B}\times\boldsymbol{J}\right] \\
{K}_\text{H}&=-\frac{1}{4}\boldsymbol{\nabla}\cdot\boldsymbol{{H}}
+\frac{1}{4} C_{\rho}\left[\boldsymbol{B}\times\boldsymbol{j},\boldsymbol{J}\right], \label{stKa}
\end{align}
are connected, similarly to the previous case, with the third order structure functions
\begin{align}
\boldsymbol{{Y}}&=\left\langle \delta\boldsymbol{u} \left(|\delta\boldsymbol{w}|^2+
 |\delta\boldsymbol{B}|^{2}\right)-2\delta\boldsymbol{B}\left(\delta\boldsymbol{u}\cdot\delta\boldsymbol{B}\right)\right\rangle, \nonumber \\
\boldsymbol{{H}}&= \left\langle \delta\boldsymbol{B}\left(\delta\boldsymbol{j}\cdot\delta\boldsymbol{B}\right)-\frac{1}{2}\delta\boldsymbol{j}|\delta\boldsymbol{B}|^{2}\right\rangle.
\end{align}
The third order structure functions $\boldsymbol{{Y}}$ and $\boldsymbol{{H}}$ represent the
compressible generalizations of $\boldsymbol{{Y}}^{(i)}$ and $\boldsymbol{{H}}^{(i)}$
{
(note that $\boldsymbol{{H}}=\langle\rho \rangle \boldsymbol{{H}}^{(i)}$), whereas $R$ is a compressible term
that does not seem to be easily expressible as a structure function but can be given as
\begin{align}
{R}=\left\langle
\delta\boldsymbol{w}\cdot \left( \theta^\prime \boldsymbol{w}-\theta \boldsymbol{w}^\prime \right)\right\rangle
\end{align}
where $\theta$ is the dilatation field, 
\begin{align}
\theta=\boldsymbol{\nabla}\cdot\boldsymbol{u},
\end{align}
} and where the primes denote quantities evaluated at $\boldsymbol{x}^\prime=\boldsymbol{x}+\boldsymbol{l}$.
Equation~(\ref{stKa}) defining 
${K}_\text{MHD}$ and ${K}_\text{H}$ also involves ``correction'' terms related to the density variations given by
\begin{align*}
C_{\rho}\left[\boldsymbol{a},\boldsymbol{b}\right]
=\left\langle\left(\frac{\rho^{\prime}}{\rho}-1\right)\boldsymbol{a}^{\prime}\cdot\boldsymbol{b}+\left(\frac{\rho}{\rho^{\prime}}-1\right)\boldsymbol{a}\cdot\boldsymbol{b}^{\prime} \right\rangle.
\end{align*}

Similarly to the incompressible Equation~(\ref{epsilonstar}), in Equation~(\ref{qstar}) 
$S_{w}$ and {$S_{B}= \langle\rho \rangle S_b^{(i)}$} represent
the separation-scale distribution of the kinetic and magnetic energy,
 ${K}_\text{MHD}$ and ${K}_\text{H}$ are the MHD and Hall
cascade rates, respectively, and $\eta \Delta {S}_B/2$ describes the resistive dissipation.
Furthermore,  $\varPsi$ represents the pressure-strain effect,
\begin{align}
\varPsi &= -\frac{1}{2}\left\langle \delta \boldsymbol{w}\cdot  \delta \left(\frac{\boldsymbol{\nabla}\cdot \boldsymbol{\mathrm{P}} }{\sqrt{\rho}}\right) \right \rangle +Q_\Sigma.
\end{align}
This term is different from that defined in \cite{hellal21b}, here we have added the pressure-strain rate $Q_\Sigma$.

Equation~(\ref{qstar}) describes an energy-budget relationship and the KHM equation
corresponds to the equality $Q^*=Q=\text{const}$.
Figure~\ref{yag} shows the test of the KHM equation, the cascade rate $Q^*$
and its contributing terms
as functions of the separation scale $l$ in the three simulations.
Figure~\ref{yag} demonstrates that the compressibility and
inclusion of the pressure-strain effect significantly improve
the conservation of the effective cascade/dissipation rate.
$Q^*$ is relatively constant in the three simulations, and 
the decrease of MHD + Hall cascade rate observed at the transition from the MHD to sub-ion scales
is compensated by the resistive and the pressure-strain coupling.
The contribution from this pressure-strain coupling term is of similar amplitude 
as the resistive one at low and moderate beta, while it dominates
in high beta case of run~3. Beside the constancy, $Q^*$ is
close to its expected value $Q$; the relative error is small
$|Q^*|/Q \lesssim 0.1$ and is likely related to numerical
limitation of the code.

Comparing more in detail individual contributions in the two approaches,
we see that the compressibility and density variations 
are not important in the current simulations.
The Hall contributions are almost the same, $\langle\rho\rangle K_\text{H}^{(i)}\simeq K_\text{H}$ 
the MHD contributions are close to each other, $\langle\rho\rangle K_\text{MHD}^{(i)} \sim K_\text{MHD}$,
the relative difference is about $10 \%$  for $\beta_i=0.1$ 
(mostly owing to the compressible term $R$) and decreases
for higher betas. The main difference between the presentations
of these terms in Figures~\ref{iyag} and~\ref{yag} is owing to
the different normalizations ($\epsilon$ vs $Q$); 
this normalization has actually a highly physical
meaning that expresses the inertial-range picture relating the
dissipation and cascade rates. In order to have the cascade rate
equal to the dissipation one, one has to include the pressure-strain contribution
to the dissipation rate.
We conclude that the substantial difference between the 
Equations~(\ref{epsilonstar}) and~(\ref{qstar}) { for
the three simulations investigated in this paper}
is owing to the inclusion of the pressure-strain effect.

\section{Discussion}
\label{discus}
In this paper we investigated the transition from MHD to
sub-ion scales in 2D hybrid simulations using
the KHM equations. 
{
We analyzed three 2D hybrid simulations with parameters similar
(but not identical) to those in \cite{hellal18}. We observe
qualitatively the same results as in \cite{hellal18}, the incompressible, constant-density
KHM equation shows that the effective cascade/dissipation rate decreases
at sub-ion scales.
Note that the incompressible Hall cascade rate in \cite{hellal18} is twice the correct
value \cite[cf.,][]{ferral19} so that the decrease of the effective incompressible
cascade rate at sub-ion scales is underestimated there.
The decrease of the incompressible rate at sub-ion scales is in agreement with other previous simulation
and observation works \citep{bandal20b,adhial21}.
}

The compressible version of the KHM equation \citep{hellal21b}
 which includes the pressure-strain coupling \citep{yangal17}
exhibits a good conservation property,
the effective cascade/dissipation rate
is constant; the decrease of the cascade rate
an ion scales is compensated by the resistive
effects, and, more importantly, the pressure-strain
coupling. The effective cascade/dissipation rate
is close to the expected value. We observe a small
but non-negligible discrepancy
that is likely connected with numerical issues 
of the particle-in-cell, finite difference scheme. 
The discrepancy between the prediction of KHM equation
and the simulation results depends on
many numerical as well as physical parameters.
In fact, the KHM equation may serve as a test of numerical codes.
We were not able to discern and analyze all the relevant
parameters.
Additional simulations show that the error decreases with an
increasing number of particles per cell and decreasing
beta. This indicates that the numerical noise due to
the limited number of particles per cell contributes
to the error; the finite-difference scheme for
the resistivity is a possible source of additional
errors, and the pressure-strain coupling
is prone to be affected by the numerical noise.

In a more general context, our study suggests that, 
for quite a wide range of plasma and turbulence parameters
the actual compressibility and density variations
are weak and one can use an incompressible
approach but pressure-strain effect needs to be retained.
The same probably applies for the weakly compressible
solar wind \citep{zankal17}.
The spatial scale of the pressure-strain onset 
increases with $\beta_i$ (and $\rho_i$). Our
current results do not, however, show a clear
connections between this onset scale
and $\rho_i$  and/or $d_i$. The characteristic
scales of the pressure-strain channel, $\boldsymbol{\mathrm{P}} : \boldsymbol{\Sigma}$,
are apparently difficult to determine \citep{depe18}.

The pressure-strain effect is in principle reversible
but, in contrast to fluid cases \citep{hellal21a,hellal21b},
it appears to work only in one direction and acts
as an effective dissipation rate in the kinetic case.
Our kinetic/hybrid KHM results are similar to 
the spatial filtering analysis of fully kinetic simulations \citep{mattal20}
and constitute a complementary and independent confirmation
that pressure-strain coupling plays the role of a dissipation channel
that appears at ion scales. 
More kinetic simulations are needed to discern its characteristic
scales and the roles of different ion and electron species and their parameters; 
the KHM equation needs to be extended to the multifluid framework \citep{andral16}.
We also note that multiple different processes may contribute
to the pressure-strain effect, such as quasi-linear damping,
magnetic reconnection, etc., and one has to go beyond
the KHM equation to distinguish them.

We plan to continue this work using 
the spectral transfer analysis that allows a clearer 
connection between the different energy-transfer channels and
the power spectra \cite[cf.,][]{papial21b}. We also plan
to extend this analysis to a three-dimensional geometry \cite[cf.,][]{verdal15,franal18a}
to investigate anisotropy of the energy cascade and 
dissipation including the pressure-strain effect.
Results obtained in this work by means of numerical simulations
needs to be complemented by in situ observations. In this respect,
 it is already challenging to extend the incompressible KHM equation to the Hall regime
\citep{bandal20b}. However, multipoint spacecraft observations (by, e.g., MMS)
may possibly be used to measure the compressible Hall MHD in situ, 
also including the pressure-strain effect
 \citep{bandal21}.

This work was performed using the Cambridge Service for Data Driven Discovery (CSD3), 
part of which is operated by the University of Cambridge Research Computing on behalf
of the STFC DiRAC HPC Facility. The DiRAC component of CSD3 was funded by BEIS capital funding via STFC capital grants ST/P002307/1 and ST/R002452/1 and STFC operations grant ST/R00689X/1.
This work also used computing resources provided by STFC DiRAC HPC Facility at Durham 
(grants ST/P002293/1, ST/R002371/1, ST/S002502/1, ST/R000832/1) for project ``dp170`` 
and by Cineca and INAF (Accordo Quadro MoU Nuove frontiere in Astrofisica) for 
project ``INA20\_C6A55''. LF is supported by the STFC grant ST/T00018X/1.


\begin{thebibliography}{}
\expandafter\ifx\csname natexlab\endcsname\relax\def\natexlab#1{#1}\fi

\bibitem[{Adhikari {et~al.}(2021)Adhikari, Parashar, Shay, Matthaeus, Pyakurel,
  Fordin, Stawarz, \& Eastwood}]{adhial21}
Adhikari, S., Parashar, T.~N., Shay, M.~A., {et~al.} 2021, Phys. Rev. E, 104,
  065206, doi:10.1103/PhysRevE.104.065206

\bibitem[{Andr{\'e}s {et~al.}(2018)Andr{\'e}s, Galtier, \& Sahraoui}]{andral18}
Andr{\'e}s, N., Galtier, S., \& Sahraoui, F. 2018, Phys. Rev. E, 97, 013204

\bibitem[{Andr{\'e}s {et~al.}(2016)Andr{\'e}s, Mininni, Dmitruk, \&
  G{\'o}mez}]{andral16}
Andr{\'e}s, N., Mininni, P.~D., Dmitruk, P., \& G{\'o}mez, D.~O. 2016, Phys.
  Rev. E, 93, 063202

\bibitem[{Bandyopadhyay {et~al.}(2020)Bandyopadhyay, Sorriso-Valvo, Chasapis,
  Hellinger, Matthaeus, Verdini, Landi, Franci, Matteini, Giles, Gershman,
  Pollock, Russell, Strangeway, Torbert, Moore, \& Burch}]{bandal20b}
Bandyopadhyay, R., Sorriso-Valvo, L., Chasapis, A., {et~al.} 2020, Phys. Rev.
  Lett., 124, 225101, doi:10.1103/PhysRevLett.124.225101

\bibitem[{{Bandyopadhyay} {et~al.}(2021){Bandyopadhyay}, {Chasapis},
  {Matthaeus}, {Parashar}, {Haggerty}, {Shay}, {Gershman}, {Giles}, \&
  {Burch}}]{bandal21}
{Bandyopadhyay}, R., {Chasapis}, A., {Matthaeus}, W.~H., {et~al.} 2021, Phys.
  Plasmas, 28, 112305, doi:10.1063/5.0071015

\bibitem[{Bruno \& Carbone(2013)}]{brca13}
Bruno, R., \& Carbone, V. 2013, LRSP, 10, 2, doi:10.12942/lrsp-2013-2

\bibitem[{{de K\'arm\'an} \& {Howarth}(1938)}]{kaho38}
{de K\'arm\'an}, T., \& {Howarth}, L. 1938, Proc. Royal Soc. London Series A,
  164, 192

\bibitem[{{Del Sarto} \& Pegoraro(2018)}]{depe18}
{Del Sarto}, D., \& Pegoraro, F. 2018, MNRAS, 475, 181,
  doi:10.1093/mnras/stx3083

\bibitem[{Ferrand {et~al.}(2019)Ferrand, Galtier, Sahraoui, Meyrand,
  {Andr\'es}, \& Banerjee}]{ferral19}
Ferrand, R., Galtier, S., Sahraoui, F., {et~al.} 2019, ApJ, 881, 50,
  doi:10.3847/1538-4357/ab2be9

\bibitem[{Franci {et~al.}(2016{\natexlab{a}})Franci, Hellinger, Matteini,
  Verdini, \& Landi}]{franal16a}
Franci, L., Hellinger, P., Matteini, L., Verdini, A., \& Landi, S.
  2016{\natexlab{a}}, in Proc. 14th Int. Solar Wind Conf., Vol. 1720 (AIP)

\bibitem[{Franci {et~al.}(2015)Franci, Landi, Matteini, Verdini, \&
  Hellinger}]{franal15b}
Franci, L., Landi, S., Matteini, L., Verdini, A., \& Hellinger, P. 2015, ApJ,
  812, 21, doi:10.1088/0004-637X/812/1/21

\bibitem[{Franci {et~al.}(2016{\natexlab{b}})Franci, Landi, Matteini, Verdini,
  \& Hellinger}]{franal16b}
---. 2016{\natexlab{b}}, ApJ, 833, 91, doi:10.3847/1538-4357/833/1/91

\bibitem[{Franci {et~al.}(2018)Franci, Landi, Verdini, Matteini, \&
  Hellinger}]{franal18a}
Franci, L., Landi, S., Verdini, A., Matteini, L., \& Hellinger, P. 2018, ApJ,
  1, 26, doi:10.3847/1538-4357/aaa3e8

\bibitem[{Frisch(1995)}]{fris95}
Frisch, U. 1995, Turbulence (Cambridge University Press)

\bibitem[{{Galtier}(2006)}]{galt06}
{Galtier}, S. 2006, J. Plasma Phys., 72, 721, doi:10.1017/S0022377806004521

\bibitem[{{Galtier}(2008)}]{galt08}
---. 2008, Phys. Rev. E, 77, 015302, doi:10.1103/PhysRevE.77.015302

\bibitem[{{Ghosh} {et~al.}(1996){Ghosh}, {Siregar}, {Roberts}, \&
  {Goldstein}}]{ghosal96}
{Ghosh}, S., {Siregar}, E., {Roberts}, D.~A., \& {Goldstein}, M.~L. 1996, J.
  Geophys Res., 101, 2493, doi:10.1029/95JA03201

\bibitem[{Hellinger {et~al.}(2021{\natexlab{a}})Hellinger, Papini, Verdini,
  Landi, Franci, Matteini, \& {Montagud-Camps}}]{hellal21b}
Hellinger, P., Papini, E., Verdini, A., {et~al.} 2021{\natexlab{a}}, ApJ, 917,
  101, doi:10.3847/1538-4357/ac088f

\bibitem[{Hellinger {et~al.}(2018)Hellinger, Verdini, Landi, Franci, \&
  Matteini}]{hellal18}
Hellinger, P., Verdini, A., Landi, S., Franci, L., \& Matteini, L. 2018, ApJL,
  857, L19, doi:10.3847/2041-8213/aabc06

\bibitem[{Hellinger {et~al.}(2021{\natexlab{b}})Hellinger, Verdini, Landi,
  Franci, Papini, \& Matteini}]{hellal21a}
Hellinger, P., Verdini, A., Landi, S., {et~al.} 2021{\natexlab{b}}, Phys. Rev.
  Fluids, 6, 044607, doi:10.1103/PhysRevFluids.6.044607

\bibitem[{{Kida} \& {Orszag}(1990)}]{kior90}
{Kida}, S., \& {Orszag}, S.~A. 1990, J. Sci. Comput., 5, 85

\bibitem[{{Leamon} {et~al.}(1998){Leamon}, {Smith}, {Ness}, {Matthaeus}, \&
  {Wong}}]{leamal98}
{Leamon}, R.~J., {Smith}, C.~W., {Ness}, N.~F., {Matthaeus}, W.~H., \& {Wong},
  H.~K. 1998, J. Geophys Res., 103, 4775, doi:10.1029/97JA03394

\bibitem[{Matteini {et~al.}(2020)Matteini, Franci, Alexandrova, Lacombe, Landi,
  Hellinger, Papini, \& Verdini}]{mattal20b}
Matteini, L., Franci, L., Alexandrova, O., {et~al.} 2020, Front. Astron. Space
  Sci., 7, 563075, doi:10.3389/fspas.2020.563075

\bibitem[{Matthaeus {et~al.}(2020)Matthaeus, Yang, Wan, Parashar,
  Bandyopadhyay, Chasapis, Pezzi, \& Valentini}]{mattal20}
Matthaeus, W.~H., Yang, Y., Wan, M., {et~al.} 2020, ApJ, 891, 101,
  doi:10.3847/1538-4357/ab6d6a

\bibitem[{Matthews(1994)}]{matt94}
Matthews, A. 1994, JCoPh, 112, 102, doi:10.1006/jcph.1994.1084

\bibitem[{{Mininni} \& {Pouquet}(2009)}]{mipo09}
{Mininni}, P.~D., \& {Pouquet}, A. 2009, Phys. Rev. E, 80, 025401,
  doi:10.1103/PhysRevE.80.025401

\bibitem[{Monin \& Yaglom(1975)}]{moya75}
Monin, A.~S., \& Yaglom, A.~M. 1975, Statistical fluid mechanics: Mechanics of
  turbulence (Cambridge)

\bibitem[{Papini {et~al.}(2019)Papini, Franci, Landi, Verdini, Matteini, \&
  Hellinger}]{papial19}
Papini, E., Franci, L., Landi, S., {et~al.} 2019, ApJ, 870, 52,
  doi:10.3847/1538-4357/aaf003

\bibitem[{Papini {et~al.}(2021)Papini, Hellinger, Verdini, Landi, Franci,
  Montagud-Camps, \& Matteini}]{papial21b}
Papini, E., Hellinger, P., Verdini, A., {et~al.} 2021, Atmosph., 12, 1632,
  doi:10.3390/atmos12121632

\bibitem[{Parashar \& Matthaeus(2016)}]{pama16}
Parashar, T.~N., \& Matthaeus, W.~H. 2016, ApJ, 832, 57,
  doi:10.3847/0004-637X/832/1/57

\bibitem[{{Pezzi} {et~al.}(2021){Pezzi}, {Liang}, {Juno}, {Cassak},
  {V{\'a}sconez}, {Sorriso-Valvo}, {Perrone}, {Servidio}, {Roytershteyn},
  {TenBarge}, \& {Matthaeus}}]{pezzal21}
{Pezzi}, O., {Liang}, H., {Juno}, J.~L., {et~al.} 2021, MNRAS, 505, 4857,
  doi:10.1093/mnras/stab1516

\bibitem[{Politano \& Pouquet(1998)}]{popo98b}
Politano, H., \& Pouquet, A. 1998, Phys. Rev. E, 57, R21

\bibitem[{Servidio {et~al.}(2015)Servidio, Valentini, Perrone, Greco, Califano,
  Matthaeus, \& Veltri}]{serval15}
Servidio, S., Valentini, F., Perrone, D., {et~al.} 2015, J. Plasma Phys., 81,
  325810107, doi:10.1017/S0022377814000841

\bibitem[{Verdini {et~al.}(2015)Verdini, Grappin, Hellinger, Landi, \&
  {M\"uller}}]{verdal15}
Verdini, A., Grappin, R., Hellinger, P., Landi, S., \& {M\"uller}, W.~C. 2015,
  ApJ, 804, 119, doi:10.1088/0004-637X/804/2/119

\bibitem[{Yang {et~al.}(2019)Yang, Wan, Matthaeus, Sorriso-Valvo, Parashar, Lu,
  Shi, \& Chen}]{yangal19}
Yang, Y., Wan, M., Matthaeus, W.~H., {et~al.} 2019, MNRAS, 482, 4933,
  doi:10.1093/mnras/sty2977

\bibitem[{Yang {et~al.}(2017)Yang, Matthaeus, Parashar, Haggerty, Roytershteyn,
  Daughton, Wan, Shi, \& Chen}]{yangal17}
Yang, Y., Matthaeus, W.~H., Parashar, T.~N., {et~al.} 2017, Phys. Plasmas, 24,
  072306, doi:10.1063/1.4990421

\bibitem[{{Yordanova} {et~al.}(2021){Yordanova}, {V{\"o}r{\"o}s},
  {Sorriso-Valvo}, {Dimmock}, \& {Kilpua}}]{yordal21}
{Yordanova}, E., {V{\"o}r{\"o}s}, Z., {Sorriso-Valvo}, L., {Dimmock}, A.~P., \&
  {Kilpua}, E. 2021, ApJ, 921, 65, doi:10.3847/1538-4357/ac1942

\bibitem[{Zank {et~al.}(2017)Zank, Adhikari, Hunana, Shiota, Bruno, \&
  Telloni}]{zankal17}
Zank, G.~P., Adhikari, L., Hunana, P., {et~al.} 2017, ApJ, 835, 147,
  doi:10.3847/1538-4357/835/2/147

\end{thebibliography}
\end{document}